\documentclass[aps,prl,twocolumn,superscriptaddress,floatfix]{revtex4-2}
\usepackage{graphicx}
\usepackage{amssymb,amsmath}
\usepackage{bm}
\usepackage{dcolumn}
\usepackage[OT1]{fontenc}
\usepackage{url}
\usepackage{mathrsfs}
\usepackage{physics}
\usepackage[driverfallback=dvipdfm]{hyperref}
\hypersetup{pdfpagemode=FullScreen,colorlinks=true,breaklinks,urlcolor=blue,linkcolor=blue,citecolor=blue}

\begin{document}

\title{Light-Cone Scaling of In-Circuit Noise in Randomized Measurements}

\author{Pan Yu}
\affiliation{College of Physics, Sichuan University, Chengdu 610065, China}

\author{Yan He}
\affiliation{College of Physics, Sichuan University, Chengdu 610065, China}

\author{Yadong Wu}
\email{yadongwu@scu.edu.cn}
\affiliation{College of Physics, Sichuan University, Chengdu 610065, China}

\date{\today}

\begin{abstract}
Randomized measurements provide an efficient way to extract physical properties of an unknown quantum state from limited data. On near-term hardware, gate and readout errors bias the reconstructed observables. Here we develop a microscopic description of this bias for locally scrambled shallow circuits. Independent local twirling reduces local implementation noise to stochastic Pauli damping, and a noise event contributes only when it overlaps the Heisenberg evolution of the measured Pauli operator. This gives an activated path-average formula for the noisy Pauli coefficient. In one-dimensional shallow circuits, the activated noise volume grows linearly with the size of a contiguous observable, leading to an exponential damping ratio. We verify this scaling for two-qubit random Clifford and locally scrambled iSWAP circuits with  two-qubit Pauli noise, including spatial fluctuations and temporal drift. The scaling supports a small-string calibration protocol that predicts larger string observables without learning the full noisy measurement channel. Our result relates the noise bias of shallow-shadow protocols directly to operator-evolving dynamics.
\end{abstract}

\maketitle

\textbf{Introduction.--}
Randomized measurements, most notably classical shadows \cite{Huang:2020aa,aaronson2018shadowtomographyquantumstates}, provide a useful protocol for estimating many properties of an unknown quantum state. By applying random unitaries and measuring in the computational basis, one can predict fidelities, many-body correlators, entropies, and other observables with substantially fewer samples than full tomography \cite{PhysRevLett.120.050406,doi:10.1126/science.aau4963,PhysRevLett.125.200501,PhysRevLett.127.030503,Hadfield:2022aa,Huang:2022aa,Elben:2023aa}. Different random ensembles lead to different reconstruction channels and sample costs, ranging from single-qubit Clifford measurements to shallow, Hamiltonian-driven, locally entangled, and locally scrambled circuits \cite{PhysRevResearch.4.013054,PhysRevResearch.5.023027,PhysRevLett.130.230403,Akhtar2023scalableflexible,zhang2024holographicclassicalshadowtomography,Bu:2024aa,Ippoliti2024classicalshadows,PhysRevLett.133.020602,Zhou2024efficientclassical,Akhtar2025dualunitaryshadow,Wu:2026aa}.

On realistic devices, the random circuit is implemented with local gates and is subject to gate and readout noise \cite{Preskill2018quantumcomputingin}. These errors modify the measurement probabilities and bias the shadow estimator. Existing noisy-shadow and mitigation methods usually represent the accumulated imperfection by an effective measurement channel, which can then be calibrated, inverted, or extrapolated in post-processing \cite{PRXQuantum.2.030348,hu2022logicalshadowtomographyefficient,Koh2022classicalshadows,PRXQuantum.4.010303,Wu:2024aa,onorati2024noisemitigatedrandomizedmeasurementsselfcalibrating,PRXQuantum.5.010324,PhysRevLett.133.130803,Hu:2025aa,PhysRevLett.134.090801}. This description is useful for robustness guarantees and practical correction, but it does not identify how local errors are accumulated during operator evolution inside the random circuit.

In this work, we give a microscopic picture for locally scrambled shallow circuits \cite{PhysRevB.101.224202,PhysRevB.102.134203,PhysRevResearch.5.023027}. We consider weak local completely positive trace-preserving (CPTP) noise that is independent of the sampled random gates, with independent local twirls between circuit layers and no leakage outside the computational Hilbert space. Under these conditions, each local implementation noise channel is randomly compiled to an effective Pauli-diagonal channel \cite{PhysRevA.59.1070,PhysRevA.94.052325,Cai:2019aa,Chen:2023aa}. In the Heisenberg picture, a local noise channel damps the signal only when it overlaps the support of the evolving Pauli operator. We call such an overlap an {\it activated} noise event. The resulting damping is controlled by the space-time history of operator evolution.

\begin{figure}
 \begin{center}
  \includegraphics[width=0.98\linewidth]{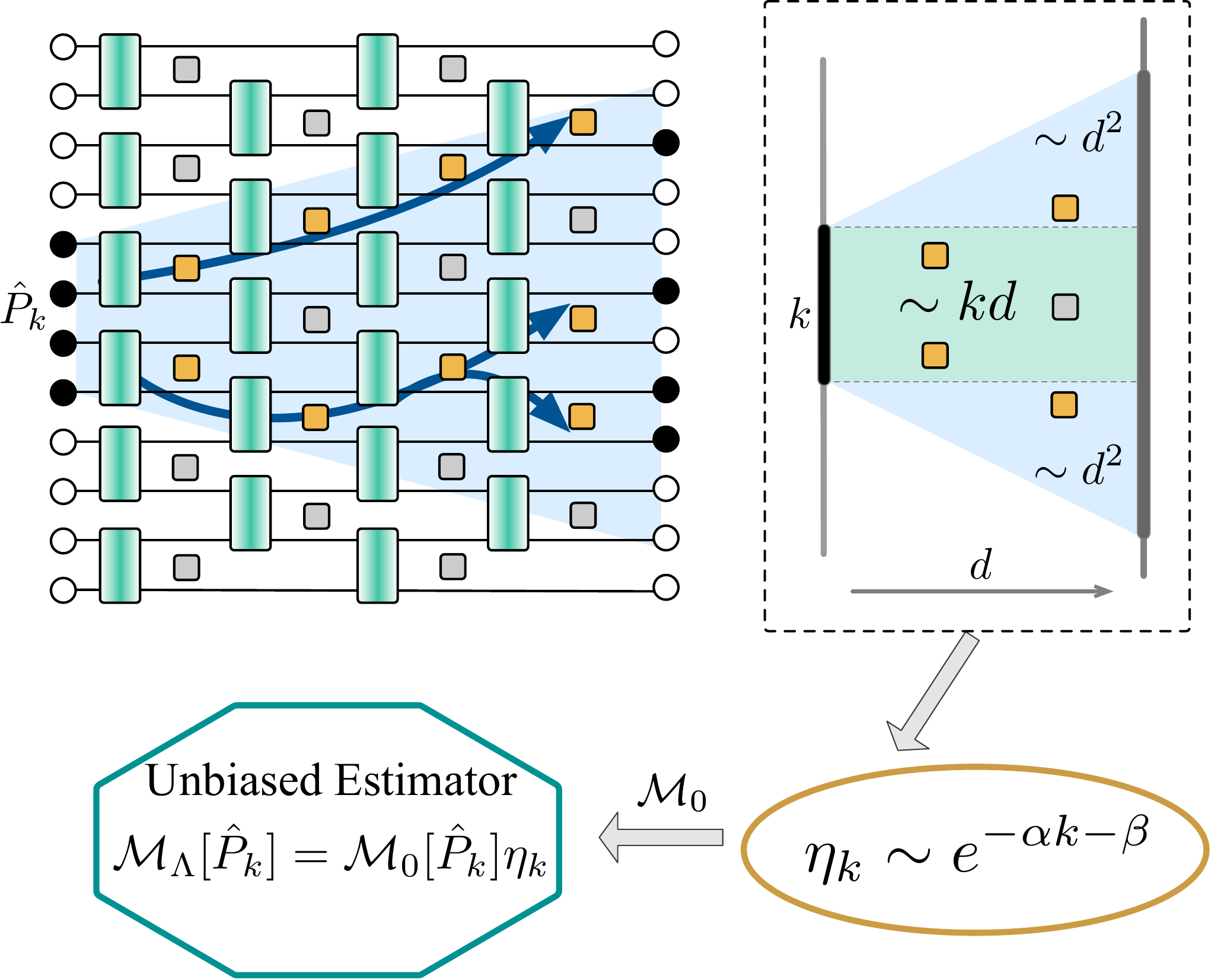}
    \caption{\textbf{Activated noise in shallow randomized measurements.} A locally scrambled shallow circuit contains local two-qubit gates and noise channels. In the Heisenberg picture, a noise channel is activated when it overlaps the evolving Pauli support, denoted as orange plaquettes, while inactive channels outside the path denoted as gray plaquettes do not damp the corresponding coefficient. For a contiguous size-$k$ operator, the light cone contains a bulk contribution of order $kd$ and two finite-depth fronts of order $d^2$. The accumulated damping therefore follows $\eta_k\sim e^{-\alpha k-\beta}$, giving the displayed relation $w_{\mathcal E_U,\Lambda}(\hat P_k)=\eta_k w_{\mathcal E_U,0}(\hat P_k)$ used to calibrate a noisy shadow estimator.}
    \label{fig:NoiseShadow}
 \end{center}
\end{figure}

In an ideal shadow protocol, the final operator-size distribution determines the shadow estimator and sample efficiency \cite{PhysRevResearch.5.023027,PhysRevLett.130.230403,Bu:2024aa}. With in-circuit noise, the bias instead depends on the full light cone of the operator. For a contiguous size-$k$ operator in a shallow one-dimensional circuit of depth $d$, the light-cone bulk is proportional to $kd$, while the boundary gives a depth-dependent contribution set by the spreading velocity \cite{PhysRevX.8.021013,PhysRevX.8.021014}. This geometry yields an exponential law in $k$ described by a slope and an offset. We derive this law from an activated path average, verify it for anisotropic and fluctuating local noise, and use it to calibrate larger string observables from simple product-state data. Thus the protocol learns a geometry-resolved damping law instead of reconstructing the complete noisy measurement map.

\textbf{Noisy shadows as a path average.--}
The shadow protocol applies a random unitary $\hat U$ drawn from an ensemble $\mathcal E_U$ and measures in the computational basis. Each shot gives a classical snapshot $\hat\sigma_{\hat U,z}=\hat U^\dagger |z\rangle\langle z|\hat U$. In the presence of a noisy circuit channel $\Lambda$, the mean snapshot defines the measurement channel
\begin{equation}
\mathcal M_{\mathcal E_U,\Lambda}[\hat\rho]=\mathbb E_{\hat U\sim\mathcal E_U}\sum_z\hat\sigma_{\hat U,z}\,
\mel{z}{\Lambda[\hat U\hat\rho\hat U^\dagger]}{z}.
\label{eq:noisy_measurement_channel}
\end{equation}
For a locally scrambled ensemble, independent local sign flips in the ensemble average remove every off-diagonal Pauli component of this channel. The ideal and noisy channels are therefore diagonal in the Pauli basis, which are denoted as 
$\mathcal M_{\mathcal E_U,0}[\hat P]=w_{\mathcal E_U,0}(\hat P)\hat P,$ and 
    $\mathcal M_{\mathcal E_U,\Lambda}[\hat P]=w_{\mathcal E_U,\Lambda}(\hat P)\hat P$.
Both channels are independent of the input state and of the Pauli labels within a fixed support geometry. The remaining coefficient depends on the support geometry of $\hat P$, its possible support histories, and the local noisy kernels intersecting these histories. The corresponding estimator is unbiased when it is normalized by $w_{\mathcal E_U,\Lambda}(\hat P)$ \cite{PhysRevResearch.5.023027,PhysRevLett.130.230403,Bu:2024aa} that 
\begin{equation}
\Tr(\hat P\hat \rho)=w_{\mathcal E_U,\Lambda}(\hat P)^{-1}\rm Tr(\hat P\mathcal M_{\mathcal E_U,\Lambda}[\hat\rho]),
\label{eq:shadowEst}
\end{equation}
leading to the corresponding sample variance as $\Vert\hat P\Vert^2=w_{\mathcal E_U,\Lambda}(\hat P)^{-2}w_{\mathcal E_U,0}(\hat P)$. Detailed derivations are given in the Supplemental Material \cite{supp}.

For shallow locally scrambled unitary ensemble,  the depth-$d$ circuit can be written as $\hat U=\hat U_d\cdots\hat U_1$. For a measured Pauli operator $\hat P$, its Heisenberg support path depending on the unitary:
\begin{equation}
 \gamma=(\hat P_{\gamma_0},\hat P_{\gamma_1},\ldots,\hat P_{\gamma_d}),
 \qquad \hat P_{\gamma_0}=\hat P,\nonumber
 \label{eq:path_definition}
\end{equation}
where $\hat P_{\gamma_\ell}$ is the Pauli support after propagation through $\ell$ local layers, denoting the set of sites on which the Pauli string is nonidentity. And $\Pr(\gamma|\hat P,\mathcal E_U)$ is the associated support-path probability. Its final support size is $m(\gamma)=|\mathrm{supp}(\hat P_{\gamma_d})|$. Thus the ideal coefficient is a path average over the ensemble-averaged Heisenberg support evolution,
\begin{equation}
    w_{\mathcal E_U,0}(\hat P)
    =\sum_\gamma
    \frac{\Pr(\gamma|\hat P,\mathcal E_U)}{3^{m(\gamma)}} .
    \label{eq:ideal_weight}
\end{equation}
 At fixed final support, local scrambling distributes the $\hat X$, $\hat Y$, and $\hat Z$ labels uniformly. Only the all-$\hat Z$ label assignment contributes to computational-basis readout, giving the factor $3^{-m(\gamma)}$. For Clifford circuits, $\gamma$ is a stochastic Pauli-string trajectory. For non-Clifford locally scrambled gates, it is the classical support-probability path generated by the ensemble-averaged Pauli second moment \cite{supp}.

We now insert local noise channels at space-time locations $a$. Since each unitary layer obeys locally scrambled symmetry, every local noise channel is twirled separately and acts as a Pauli damping factor on the local Pauli component. Along a path $\gamma$, define $I_a(\gamma)=1$ when the noisy location overlaps the instantaneous Pauli support and $I_a(\gamma)=0$ otherwise. When a noise channel is activated, $\lambda_a(\gamma)$ is the damping rate depending on the local operator supports. Independent local twirls give the factor $e^{-\lambda_a(\gamma)}$. The noisy coefficient is consequently
\begin{equation}
    w_{\mathcal E_U,\Lambda}(\hat P)
    =\sum_\gamma
      \frac{\Pr(\gamma|\hat P,\mathcal E_U)}{3^{m(\gamma)}}
      \exp\!\left[-\sum_a I_a(\gamma)\lambda_a(\gamma)\right].
    \label{eq:noisy_weight}
\end{equation}

Dividing by Eq.~(\ref{eq:ideal_weight}) defines the measurement-weighted path measure $\nu(\gamma|\hat P)=\Pr(\gamma|\hat P,\mathcal E_U)3^{-m(\gamma)}/w_{\mathcal E_U,0}(\hat P)$ and the damping ratio
\begin{align}
    \eta_{\mathcal E_U,\Lambda}(\hat P)
    &\equiv
    \frac{w_{\mathcal E_U,\Lambda}(\hat P)}{w_{\mathcal E_U,0}(\hat P)}
    =\mathbb E_{\nu,\lambda}e^{-X(\hat P)},
    \label{eq:eta_path}
\end{align}
where $X(\hat P)=\sum_a I_a(\gamma)\lambda_a(\gamma)$. The measure $\nu$ is not the bare support distribution but favors paths with smaller final support through the factor $3^{-m(\gamma)}$. Thus the in-circuit noise is averaged over the operator histories that contribute to the Pauli coefficient. While end-circuit readout noise is a boundary-noise case of this formula, whereas global depolarizing noise gives support-independent damping \cite{supp}.

\textbf{Light-cone volume law.--}
For a contiguous Pauli string $\hat P_k$, let $X_k=X(\hat P_k)$ and $\eta_k= \eta_{\mathcal E_U,\Lambda}(\hat P_k)$. The joint average in Eq.~(\ref{eq:eta_path}) is over the measurement-weighted path distribution and the hardware-rate distribution. Its cumulant expansion gives
\begin{equation}
    \log\eta_k=\sum_{n\geq1}\frac{(-1)^n}{n!}\kappa_n(X_k).
    \label{eq:cumulant_expansion}
\end{equation}
For a contiguous string operator $\hat P_k$ away from the system boundary, the backward evolution is confined to a causal light cone. Its bulk contains $O(kd)$ gates, while its two fronts contain $O(d^2)$ gates \cite{PhysRevX.8.021013,PhysRevX.8.021014}. For a shallow circuit with local two-qubit noise channels following the local gates, each cumulant has the form $\kappa_n(X_k)=k\,a_n(d)+b_n(d)$. The coefficient $a_n(d)$ denotes the connected contributions to the $n$th cumulant from the bulk, whereas $b_n(d)$ collects contributions from the two light-cone fronts and other finite-depth boundary terms. Consequently,
\begin{align}
    \log\eta_k&=-\alpha(d,\Lambda)k-\beta(d,\Lambda),
    \label{eq:volume_law}\\
    \alpha(d,\Lambda)&=\sum_{n\geq1}\frac{(-1)^{n+1}}{n!}a_n(d),
    \nonumber\\
    \beta(d,\Lambda)&=\sum_{n\geq1}\frac{(-1)^{n+1}}{n!}b_n(d).
    \label{eq:alpha_beta_cumulants}
\end{align}
Thus $\alpha$ and $\beta$ are sums of connected-cumulant contributions. For locally scrambling shallow dynamics and weak local noise, the cumulant expansion is often dominated by its first cumulant. Eq.~(\ref{eq:volume_law}) applies at a fixed shallow depth $d$ while the operator must be placed away from the physical boundary and size-$k$ needs to be longer than an ensemble-dependent crossover scale. Both gate-independent Gaussian fluctuations and deterministic temporal-drift noise preserve this structure. In contrast, a random fluctuation shared by every gate in a layer can correlate an entire light-cone slice and generate nonlinear corrections in $k$ \cite{supp}.

\begin{figure}
 \begin{center}
  \includegraphics[width=0.98\linewidth]{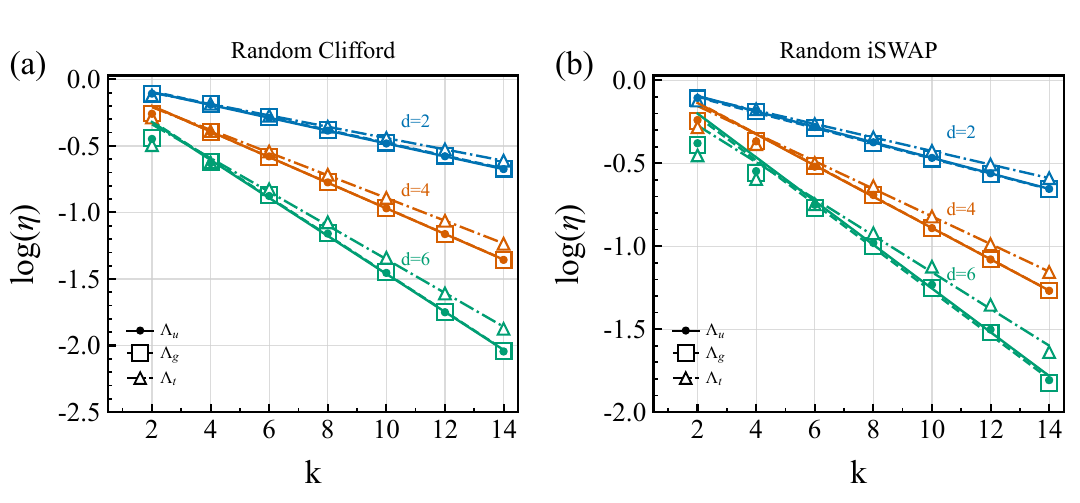}
    \caption{\textbf{Damping ratio for anisotropic and fluctuating in-circuit noise.} Transfer-matrix simulations for $N=30$ qubits with two-qubit Pauli damping rates $\lambda_{\rm L}=0.06$, $\lambda_{\rm R}=0.02$, and $\lambda_{\rm LR}=0.07$ for the $PI$, $IP$, and $PP$ sectors, respectively. We compare uniform anisotropic noise $\Lambda_u=(\lambda_{\rm L},\lambda_{\rm R},\lambda_{\rm LR})$, spatially fluctuating rates drawn from a narrow Gaussian centered at $\Lambda_u$ with width $\Lambda_u/5$, and a deterministic linear temporal drift $\Lambda_t=2t\Lambda_u/(d-1)$ for $t=0,\ldots,d-1$, keeping the same layer-averaged rate as $\Lambda_u$. Markers distinguish the three noise models and same-colored straight lines are linear fits. Colors denote depths $d=2,4,6$. The fits use $k\in[4,14]$ for random Clifford gates and $k\in[6,14]$ for locally scrambled iSWAP gates.}
    \label{fig2:diffNoise}
 \end{center}
\end{figure}

We evaluate Eq.~(\ref{eq:noisy_weight}) using the support transfer-matrix method for one-dimensional brick-wall circuits \cite{PhysRevB.101.224202}. Fig.~\ref{fig2:diffNoise} considers anisotropic two-qubit Pauli damping, in which the $PI$, $IP$, and $PP$ sectors have different rates. We consider an $N=30$-qubit system, with a contiguous size-$k$ Pauli operator $\hat P_k$ supported from the $9_{\rm th}$ qubit to the $(8+k)_{\rm th}$ qubit. For the shallow depths considered here, this choice ensures that the light cone does not reach the circuit boundary. Different operator-spreading dynamics lead to different linear windows of size-$k$ for random Clifford and locally scrambled iSWAP circuits. We therefore fit the even-$k$ branch of $\log\eta$ starting from $k=4$ for the two-qubit random Clifford circuit and from $k=6$ for the locally scrambled iSWAP circuit, with $R^2\in(0.99,0.99999)$. The linear behavior persists for both gate ensembles and for uniform rates $\Lambda_u$, spatially Gaussian-fluctuating rates $\Lambda_g$, and normalized temporally drifting rates $\Lambda_t$. The spatial fluctuations self-average over the light-cone bulk, so their results nearly overlap with the uniform case. The temporal drift is deterministic and changes the layer-resolved damping sampled by the support paths. Details of the transfer-matrix calculation and the anisotropic-rate analysis are given in the Supplemental Material \cite{supp}.

\begin{figure}
 \begin{center}
  \includegraphics[width=0.98\linewidth]{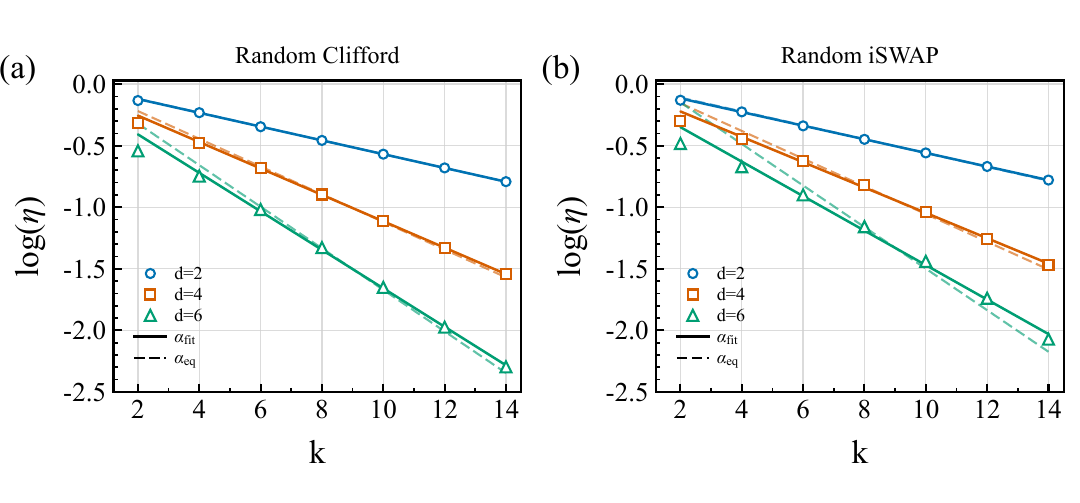}
    \caption{\textbf{Comparison with a heuristic equilibrium estimate.} Damping ratios for uniform two-qubit depolarizing noise with $\lambda_u=0.06$ at depths $d=2,4,6$. Markers show numerical data from the even-$k$ branch. Solid lines fit both $\alpha$ and $\beta$ in Eq.~(\ref{eq:volume_law}); dashed lines fix the heuristic first-cumulant equilibrium slope $\alpha_{\rm eq}=15d\lambda_u/32$ and fit only the intercept. Both fits use the even-$k$ data from $k=4$ through $k=14$; the $k=2$ points are shown but excluded from the fits. Panels (a) and (b) show random Clifford and locally scrambled iSWAP circuits, respectively.}
    \label{fig3:WeightSimu}
 \end{center}
\end{figure}

\begin{figure*}[t]
 \begin{center}
  \includegraphics[width=0.98\linewidth]{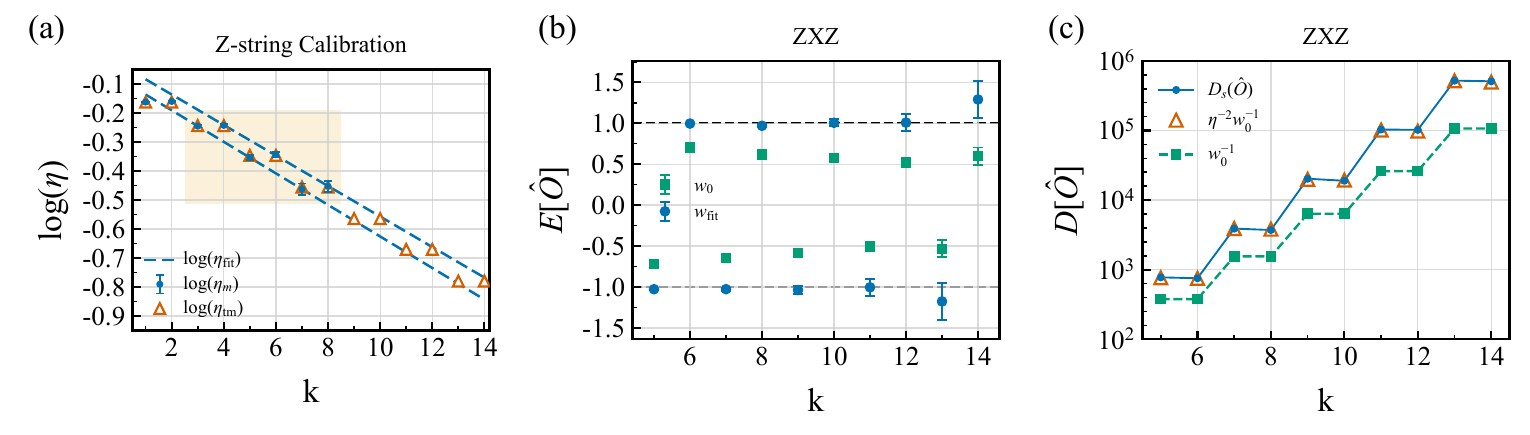}
    \caption{\textbf{Small-string calibration and prediction of larger observables.} (a) Contiguous $\hat Z_k$ strings measured on the product state $\hat\rho_0$ determine the geometry-resolved damping ratio. Small-$k$ data in the shaded fitting window determine the parity-resolved dashed fits, which extrapolate to larger $k$ and agree with the transfer-matrix results shown as triangles. (b) The calibrated coefficient corrects noisy-shadow estimates of the cluster-state string $\hat O_{ZXZ}=\hat Z_1\hat Y_2\hat X_3\cdots\hat X_{k-2}\hat Y_{k-1}\hat Z_k$, whose exact expectation value is $(-1)^k$. The calibrated estimates, $w_{\rm fit}^{-1}\mathbb{E}_U\sum_z p_\Lambda(z|U,\rho)\Tr(\hat O\hat \sigma_{U,z})$,  agree with the benchmark, whereas the uncorrected noisy estimates, $w_0^{-1}\mathbb{E}_U\sum_z p_\Lambda(z|U,\rho)\Tr(\hat O\hat\sigma_{U,z})$, shown as green squares, remain biased. (c) The measured variance $D_s[\hat O]=w_{\rm fit}^{-2}\mathbb{E}_U\sum_z p_\Lambda(z|U,\rho)\left[\Tr(\hat O\hat \sigma_{U,z})\right]^2-\langle\hat O\rangle^2$ follows the noisy shadow-norm prediction $\eta_{\rm fit}^{-2}w_0^{-1}$ and larger than the ideal value $w_0^{-1}$ because noise suppresses the measurement signal.}
    \label{fig4:Cali}
 \end{center}
\end{figure*}

The slope $\alpha$ contains information about both the local stochastic noise and the operator-spreading dynamics in the random circuit. For weak uniform noise, a simple first-cumulant reference can be obtained from the bare equilibrium support distribution. For random two-qubit Clifford gates and locally scrambled iSWAP gates, the probability that a single site carries a nonidentity Pauli operator approaches $3/4$. A two-qubit gate is therefore activated with probability $15/16$. Since the bulk light cone contains approximately $kd/2$ two-qubit gates, this gives the heuristic estimate $\alpha_{\rm eq}(d,\lambda_u)\simeq 15d\lambda_u/32$ \cite{supp}.

Fig.~\ref{fig3:WeightSimu} compares this equilibrium estimation with the numerical results for uniform two-qubit depolarizing noise with $\lambda_u=0.06$. We again calculate the Pauli-coefficient damping numerically. The solid lines are linear fits in the appropriate $k$ regime, chosen to give $R^2>0.996$. The dashed lines use the fixed slope $\alpha_{\rm eq}$ and fit only the offset, which do not reproduce the finite-depth fitted slopes. This difference shows that the linear scaling is not determined solely by the bare equilibrium support distribution. There are two competing effects. First, Jensen's inequality reduces the positive slope extracted from $\log\mathbb{E}_{\nu}e^{-X}$ relative to the first cumulant evaluated with the same measurement-weighted path distribution $\nu$. Second, at finite shallow depth, the support activation has not fully relaxed to the bare equilibrium distribution. Their competition makes $\alpha_{\rm eq}$ a heuristic benchmark \cite{supp}.

\textbf{Calibration from small strings.--}
The exponential law suggests a direct calibration protocol. One first chooses a set of observables matching same light-cone geometry and prepares a calibration state with known expectation values. For the product state $\hat\rho_0=|0\rangle\langle0|^{\otimes N}$, the ideal expectation value of every contiguous $\hat Z_k$ string is one. From eq.~\eqref{eq:shadowEst} the measured shadow signal therefore gives 
\begin{equation}
\Tr(\hat Z_k\mathcal{M}_{\mathcal{E}_U,\Lambda}[\hat\rho_0])=w_{\mathcal E_U,\Lambda}(\hat Z_k).\end{equation}
Combining it with the noiseless coefficient $w_{\mathcal E_U,0}(\hat Z_k)$ calculated from the transfer matrix gives the damping ratio $\eta_k$, from which the independent odd- and even-branch parameters $\alpha_p$ and $\beta_p$, with $p\in\{\mathrm{odd},\mathrm{even}\}$, are fitted in a small-$k$ shaded window. The fitted relation $w_{\mathcal E_U,\Lambda}(\hat P_k)\simeq e^{-\alpha_p k-\beta_p}w_{\mathcal E_U,0}(\hat P_k)$ then predicts the coefficient of larger strings in the same geometry class.

Fig.~\ref{fig4:Cali}(a) uses $10^7$ samples with a circuit-global two-qubit depolarizing probability $p_{\mathrm{circuit}}\sim\mathcal N(0.03,0.005^2)$ to extract the damping law from the calibration strings $\hat Z_9\cdots \hat Z_{8+k}$, which have the same light-cone geometry as the target observable. The scaling parameters fitted at small $k$ accurately extrapolate the damping factor to larger operator sizes. We then apply the fitted damping law to the cluster-state string $\hat O_{ZXZ} =\hat Z_9\hat Y_{10}\hat X_{11}\cdots \hat X_{6+k}\hat Y_{7+k}\hat Z_{8+k}$, 
whose exact expectation value in the ZXZ state is $\langle\hat O_{ZXZ}\rangle=(-1)^k$. As shown in Fig.~\ref{fig4:Cali}(b), the parameters learned from small $\hat Z$ strings correct the biased estimates of larger $\hat O_{ZXZ}$ strings. This transfer follows from the input-state independence of the measurement channel and its Pauli-label independence within a matched support geometry. The corresponding increase in sample variance is determined by the noisy shadow norm, $\|\hat P_k\|^2_{\mathcal E_U,\Lambda}=\eta_k^{-2}w_{\mathcal E_U,0}(\hat P_k)$, as shown in Fig.~\ref{fig4:Cali}(c). Therefore, the protocol does not learn the full noisy measurement channel. Instead, within the weak local-noise regime, it learns the geometry-resolved damping parameters required for observables with the same light-cone geometry.

\textbf{Discussion.--}
We have developed a path-based description of local in-circuit noise in randomized measurements. Local scrambling reduces local implementation noise to stochastic Pauli damping, and the damping is activated only inside the Heisenberg light cone of the measured operator. For shallow circuits, the connected-cumulant expansion gives a light-cone volume law and an exponential damping ratio for contiguous Pauli observables. The resulting slope and offset can be learned from small strings and used to calibrate larger strings with matched support geometry. For noncontiguous observables and higher-dimensional circuits, the damping analysis does not violate the path-average principle, but their volume is different from that of a contiguous string with the same Pauli weight. For example, well-separated single-site operators generate almost independent single-site cones, while nearby separated blocks can have partially overlapping cones. They should therefore be calibrated using reference observables with the same light-cone geometry rather than using the contiguous-string parameters.

This is complementary to general robust-shadow, channel-inversion, and self-calibrating mitigation protocols \cite{PRXQuantum.2.030348,hu2022logicalshadowtomographyefficient,Koh2022classicalshadows,PRXQuantum.4.010303,Wu:2024aa,onorati2024noisemitigatedrandomizedmeasurementsselfcalibrating,PRXQuantum.5.010324,PhysRevLett.133.130803,Hu:2025aa,PhysRevLett.134.090801}. Rather than treating the noisy circuit only through a global effective map, our analysis resolves which local noises contribute to a given coefficient. This protocol reduces the calibration of a fixed observable geometry to a small number of damping parameters rather than a reconstruction of the full noisy map. While random common-mode noise extends across the full cone but does not scale with the operator size, as in the case of global depolarizing noise, the damping does not follow the simple linear form. These conditions specify the intended regime of the protocol without changing its microscopic interpretation. And the noisy damping analysis also provides a direct connection between operator spreading and noise exposure \cite{doi:10.1126/science.abg5029,PhysRevA.106.012441,PhysRevLett.129.050602,PhysRevResearch.4.043141}.

\vspace{5pt}
\textit{Acknowledgement.} 
We thank Juan Yao, Hai Wang, and Xue Chen for their helpful discussions.  This project is supported by the NSFC under grants No.12504309, the Sichuan Science and Technology Program grant No.2026NSFSC0809, the Ministry of Education Key Laboratory of Quantum Physics and Photonic Quantum Information grant No.ZYGX2025K021, and the Shanghai Committee of Science and Technology grant No.25LZ2600800.

\bibliography{refNoiseShadowLC_V7}

\end{document}